\documentclass[aps,prc,reprint,twocolumn,nofootinbib,notitlepage,preprintnumbers,superscriptaddress]{revtex4-2}

\usepackage{graphicx}
\usepackage[dvipsnames]{xcolor}
\usepackage[colorlinks,allcolors=blue!80!black]{hyperref}
\usepackage{amsmath,amsthm,amssymb}
\usepackage{comment}
\usepackage{enumerate}
\newcounter{todos}
\DeclareRobustCommand{\todo}{\colorbox{pink}{\textsc{Todo}}\stepcounter{todos}}

\AtEndDocument{\typeout{}\typeout{TODO remaining: \arabic{todos}}\typeout{}}

\renewcommand{\Im}{\mathrm{Im}\,}

\newtheorem*{theorem*}{Theorem}

\begin{document}
\preprint{INT-PUB-26-017}
\preprint{LA-UR-26-24074}

\title{Light nuclear scattering from neural quantum states}

\author{Scott Lawrence}
\email{srlawrence@lanl.gov}
\affiliation{Theoretical Division, Los Alamos National Laboratory, Los Alamos, NM 87545, USA}
\author{Yukari Yamauchi}
\email{yyamauchi@lanl.gov}
\affiliation{Theoretical Division, Los Alamos National Laboratory, Los Alamos, NM 87545, USA}
\affiliation{Institute for Nuclear Theory, University of Washington, Seattle, WA 98195, USA}

\date{\today}

\begin{abstract}
	We present a method of studying few-body nuclear scattering by means of neural quantum states, without requiring time-evolution. A recently developed family of stable minimum principles for Schr\"odinger's equation provides conservative uncertainties on cross sections and partial wave amplitudes computed in this way~\cite{Lawrence:2026wrf}. We use this method to study both elastic and inelastic neutron-deuteron scattering with realistic nuclear two-body forces.
\end{abstract}

\maketitle

\section{Introduction}\label{sec:introduction}
A central object of study in nuclear and particle physics is the $S$-matrix---that unitary operator which specifies the outcome of scattering events. In weakly coupled theories such as QED, the $S$-matrix is easily computed in perturbation theory, the source of the most precise predictions in modern science~\cite{Fan:2022eto}. At any coupling, two-body scattering can be efficiently studied numerically, and these calculations have allowed nucleon-nucleon scattering experiments to be translated into knowledge of the nucleon two-body force~\cite{Stoks:1994wp,Wiringa:1994wb}. In strongly coupled many-body systems, there are few options for computing the $S$-matrix from first principles, and none that are practical for general processes with more than four dynamical particles.

Specifically in the context of nuclear dynamics, the scattering data is vital for constraining the nuclear Hamiltonian. The \emph{form} of the nuclear Hamiltonian can in principle be inferred from general effective-field-theory arguments~\cite{Somasundaram:2023sup,Cirigliano:2024ocg}, leaving a large number of unknown coefficients to be determined by fits. The most constraining data used today are the two-nucleon scattering phase shifts~\cite{Stoks:1994wp,Wiringa:1994wb}. Higher-body forces may be constrained by fitting to lattice QCD~\cite{Dawid:2025zxc,Dawid:2025doq,Contessi:2017rww}, or by using data from cold and dense systems, either bound nuclei~\cite{Curry:2025pna} or astrophysical observations~\cite{Armstrong:2026aap,Somasundaram:2024ykk}.

One approach to computing the $S$-matrix is to simulate a scattering process by evolving explicitly in time. One begins with a quantum state consisting of two well-separated ingoing wavepackets, performs time evolution, and then inspects the resulting state to determine what the species and momenta of the outgoing particles are. This approach is taken in many proposals for simulating scattering on a quantum processor~\cite{Jordan:2011ci,Jordan:2012xnu,Jordan:2014tma,Weiss:2024mie,Stetcu:2025mjw,Rule:2026brk}, although in the context of field theory these simulations are believed to be prohibitively expensive (even assuming the existence of a scalable, fault-tolerant quantum computer). Similarly, in spite of recent advances on the classical simulation of quantum time-dynamics~\cite{Alexandru:2016gsd,Alexandru:2017lqr,carleo2017solving,Lawrence:2024ebc,Lawrence:2024mnj,Lawrence:2021izu}, the calculations required for nonperturbative few-body scattering of five or more particles remain far out of reach.

In this work we propose a method of studying scattering without needing time-evolution. Scattering states---eigenstates of the Hamiltonian satisfying particular boundary conditions---have asymptotic behavior from which we may extract S-matrix elements, and therefore both partial wave amplitudes and cross sections. We find these scattering states by training a deep neural network to minimize the violation of Schr\"odinger's equation; this defines a \emph{neural scattering state} (NSS). This approach builds on recent successes in using neural quantum states in a similar manner to predict the properties of nuclear bound states~\cite{Adams:2020aax,Gnech:2021wfn,Parnes:2025seu} and responses~\cite{Parnes:2025seu}. Our approach takes as input only a Hamiltonian; neural networks have previously been trained on pre-computed exact (or nearly exact) scattering states to construct emulators capable of interpolation between Hamiltonian parameters~\cite{Zhang:2021jmi}.

Machine learning methods like the one proposed here do not come naturally equipped with reliable error estimates. In order to obtain credible uncertainties in the computed scattering properties, we exploit a particular minimum principle connecting the $L_1$ norm of the violation of Schr\"odinger's equation to a bound on the potential error in the asymptotic behavior of the scattering state~\cite{Lawrence:2026wrf}. This both informs our choice of loss function when training the neural scattering state, and provides conservative error estimates once training is complete.

The formalism of neural scattering states is detailed in Section~\ref{sec:nss}. In that section we demonstrate the method in the simplest case: a single particle, with no internal degrees of freedom, scattering from a central potential. This is extended slightly, to two-body nuclear scattering with realistic forces, in Section~\ref{sec:np}.

Two-body nuclear scattering is easily (and more efficiently) studied by other means, in particular by the radial Schr\"odinger equation. The virtue of neural scattering states lies in the fact that they generalize to few-body processes, and in particular to inelastic scattering, with two ingoing bound states and more than two outgoing states. We generalise the NSS formalism to such scattering problems in Section~\ref{sec:nd}, and demonstrate by the computation of the neutron-deuteron cross section, again from realistic two-body forces. Such calculations lie near the boundary of what is currently achievable from first principles, and are otherwise accessible by the Faddeev equations~\cite{faddeev2013quantum} which have been used to study both elastic and inelastic neutron-deuteron scattering~\cite{Phillips:1966zza,PhysRevLett.36.1438,Suslov:2013zka}. Alternative \textit{ab initio} approaches to light nuclear reaction problems include Faddeev-Yakubovsky equations~\cite{Yakubovsky:1966ue,Lazauskas:2019rfb,PhysRevC.86.044002,PhysRevC.97.044002}, quantum Monte Carlo~\cite{CARLSON198447,PhysRevC.36.27,PhysRevLett.99.022502,PhysRevLett.116.062501,PhysRevC.108.034001,Flores:2025kcr}, shell model~\cite{Fossez:2015qxa,Navratil:2010jn,Vorabbi:2026jzq,Zhang:2020rhz}, and the hyperspherical harmonic method~\cite{Kievsky_2008, PhysRevC.102.034007}. We conclude in Section~\ref{sec:discussion} with some open problems.

All simulations presented in this work make use of the python packages \texttt{jax}~\cite{deepmind2020jax} and \texttt{equinox}~\cite{kidger2021equinox}. Source code is available at~\cite{chineura}.

\section{Neural scattering states}\label{sec:nss}

We begin by describing the formalism of neural scattering states, and their use in computing both $S$-matrix elements and cross sections. In this section we will demonstrate on one-body scattering from a central potential, and in later sections we will generalize to realistic nuclear two- and three-body processes.

Let $V(x)$ be any potential on $\mathbb R^3$ (not necessarily spherically symmetric) which decays exponentially with $|x|$. We will consider the scattering of a particle of mass $M$. The system is described by the Hamiltonian
\begin{equation}\label{eq:H1}
	\hat H_1 = -\frac{\hat \nabla^2}{2M} + V(\hat x)
	\text.
\end{equation}
A scattering event can be understood in either a time-dependent or a time-independent picture. The time-dependent picture begins with a wavepacket, of small momentum uncertainty, which begins far from the origin. Under time-evolution, this wavepacket scatters off of the potential $\hat V$, and then propagates away from the origin. The $S$-matrix relates the amplitude of the ingoing wavepacket to the amplitudes of the various outgoing wavepackets. For example, for a particle with initial momentum $k$, the differential probability of observing a post-scattering momentum $p$ is given by $|\langle \vec p | \hat S | \vec k\rangle|^2$.

By taking appropriate limits we arrive at a time-independent picture, which is often more useful computationally. A scattering state $\psi$ is an eigenstate of $\hat H_1$, which represents the steady-state of the above process in the limit where the initial wavepacket is broad. A scattering state is defined by particular behavior at large radii $r$: there is only a single ingoing plane wave, of amplitude $1$ and fixed momentum $k$. The outgoing waves are uniquely determined by the requirement that the state be an eigenstate, and their amplitudes specify the $S$-matrix elements $\langle \vec p | \hat S|\vec k\rangle$ for all momenta $p$.

It is convenient to decompose a scattering state into three terms:
\begin{equation}\label{eq:decomp}
\psi = \psi_{\mathrm{in}} + \psi_{\mathrm{bulk}} + \psi_{\mathrm{out}}\text.
\end{equation}
The first term contains the only ingoing probability flux, and as such will uniquely determine the $S$-matrix elements being probed. In this work we will take $\psi_{\mathrm{in}} = e^{ikx}$ to be a conventionally normalized plane wave. The second term decays sufficiently rapidly that it may be ignored when discussing the asymptotic behavior as $r\rightarrow\infty$. The third term is responsible for all outgoing probability flux, and may be written
\begin{equation}
\psi_{\mathrm{out}} =  f(\theta, \phi) \frac{e^{i k r}}{r} + O(r^{-2})\text.
\end{equation}
The ingoing plane wave is taken to be along the $x$-direction, the angle $\theta$ is the polar angle relative to that axis, and the angle $\phi$ is the azimuthal angle. The asymptotic function $f$ encodes the $S$-matrix according to
\begin{equation}
	f(\theta,\phi) = \langle k,\theta,\phi | \hat S | k,0,0\rangle
	\text.
\end{equation}
From $f$ we can also extract cross sections and (in the case of a spherically symmetric potential $V$) partial-wave amplitudes. The total cross section is given by
\begin{equation}
	\sigma = \int \sin\theta\,d\theta\,d\phi\, |f(\theta,\phi)|^2
	\text.
\end{equation}

Thus, to determine elements of the $S$-matrix---and in particular to determine the long-time fate of an ingoing particle with momentum $\vec k$---we need only find the corresponding scattering state. Of course we cannot hope to numerically determine an exact scattering state. Some approximation must be made.

We will train a neural network to represent a state $\tilde\psi$ which has the desired ingoing wave $e^{ikx}$, but is not an exact eigenstate of the Hamiltonian $\hat H_1$. This inexactness will introduce some error into the $S$-matrix elements ineferred from the outgoing wavefunction $\tilde\psi_{\mathrm{out}}$. The details of the training procedure must be chosen to ensure that as the violation of Schr\"odinger's equation is reduced, the asymptotic behavior of $\tilde\psi$ converges to the true asymptotic behavior.

Motivated by recent results~\cite{Lawrence:2026wrf}, we define the violation of Schr\"odinger's equation by the $L_1$ norm. Specifically we write a loss functional
\begin{equation}\label{eq:L1norm_ge}
	\mathcal{L}[\tilde\psi] = \int d^3 x\, |( E-\hat H ) \tilde\psi |\,
	\equiv
	\|(E-\hat H)\tilde\psi\|_1
\end{equation}
which we will minimize to train the neural networks that constitute $\tilde\psi$. This minimization is performed over a class of ansatz wavefunction $\tilde\psi$ of the form given in Eq.~(\ref{eq:decomp}), and in particular requiring that the only ingoing wave is the plane wave $e^{ikx}$. This ensures that the only minimum is the exact scattering eigenstate.

We need the error in the asymptotic behavior to be bounded as a function of the $L_1$ loss $\mathcal{L}$. This was proven in~\cite{Lawrence:2026wrf}. Informally, the results of~\cite{Lawrence:2026wrf} may be summarized as follows.
\begin{theorem*}
	Let $\tilde\psi$ be an approximate scattering state, with ingoing behavior $e^{ikx}$. Then the error in the outgoing behavior of $\tilde\psi$ is less than or equal to $C \|(E-\hat H)\tilde\psi\|_1$, for constant $C$ determined by the potential and ingoing momentum.
\end{theorem*}
Determining the constant $C$ is often difficult, but it can be estimated empirically by looking at the behavior of $\mathcal L$ during training.

A general workflow for the NSS method is as follows. We first fix all properties of the incoming particles (potentially including bound states). Then we construct an ansatz for the scattering state $\tilde\psi$, which holds fixed the correct ingoing behavior, and allows for all possible outgoing channels. The neural networks that make up this ansatz are trained by minimizing the loss function of Eq.~(\ref{eq:L1norm_ge}). The outcome of the training is the scattering state and associated loss value. From the scattering state, we compute the observables of interest: typically cross sections or partial-wave amplitudes. The error esimates on these observables are given by observing the behavior of the loss function.

\begin{figure*}
	\centerline{
	\includegraphics[width=0.45\linewidth]{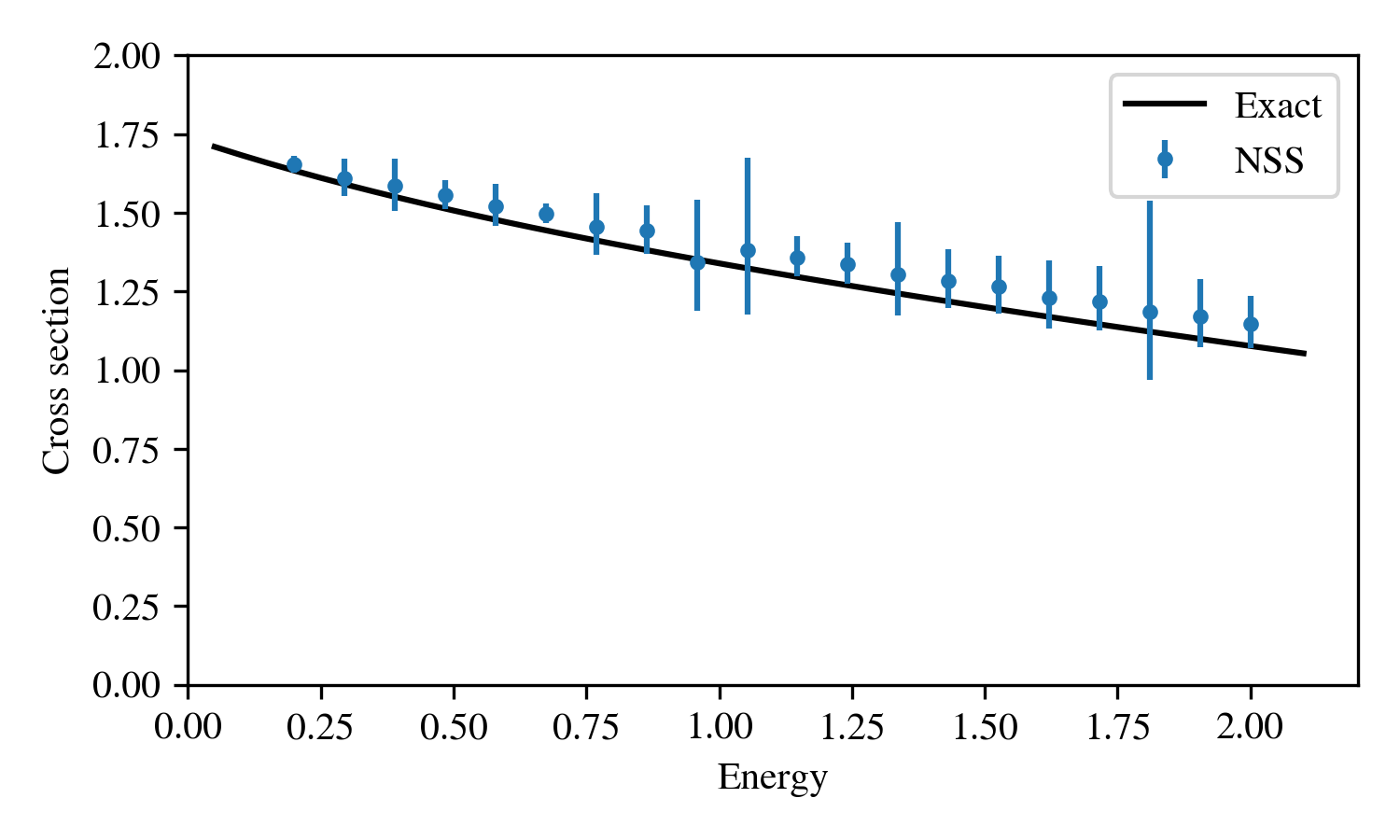}
	\hfil
	\includegraphics[width=0.45\linewidth]{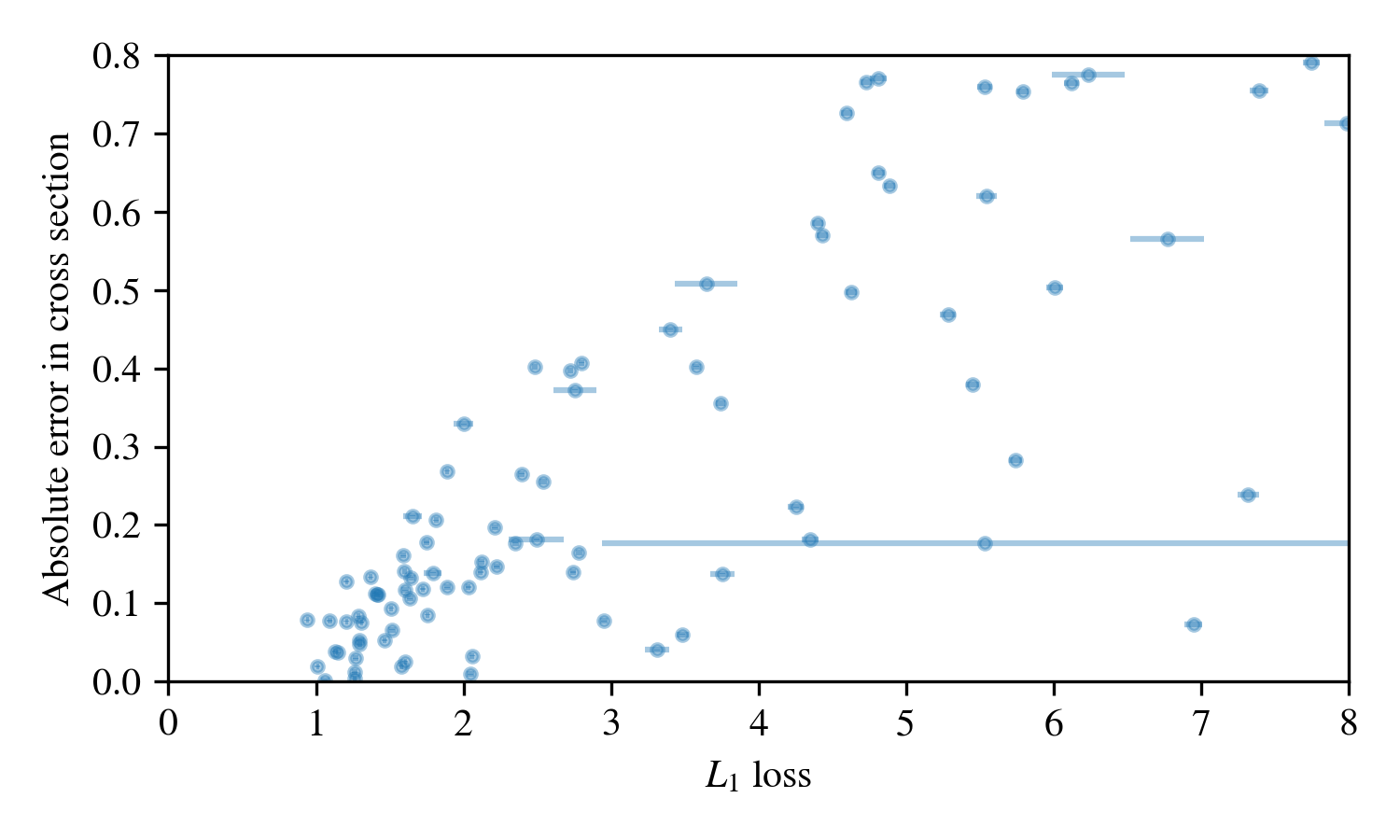}
}
	\caption{Simulation of single-particle scattering from a hard sphere, via neural scattering states. At left, the cross section is computed across a range of incident energies, for a potential with $V_0=1$ and $r_0 = 1$, and a particle of mass $M=1$. The exact cross sections are obtained by solving the radial Schr\"odinger equation. At right, the $L_1$ loss defined by Eq.~(\ref{eq:L1norm_ge}) is compared to the error in the computed cross section, for approximate wavefunctions obtained over the course of training. Evidence of the linear bound is clearly visible, and the coefficient can be estimated from such a plot. Horizontal bars indicate estimates of the statistical errors in the loss.}
	\label{fig:sphere}
\end{figure*}

In the rest of this section we demonstrate the NSS method by computing the scattering of a single particle off of the hard-sphere potential defined by
\begin{equation}
	V(r) = V_0\,\theta(r_0-r)\text.
\end{equation}
Here $V_0$ and $r_0$ are respectively the strength and range of the potential. The Heaviside step function is $\theta(\cdot)$. 
The wavefunction $\tilde\psi$ is constrained to be of the form of Eq.~(\ref{eq:decomp}); that is,
\begin{equation}
	\tilde\psi(x) = e^{ikx} + \tilde\psi_{\mathrm{bulk}}(x) + \tilde\psi_{\mathrm{out}}(x)\text.
\end{equation}
Both the outgoing and bulk wavefunctions are ultimately parameterized by neural networks. The full details of the architecture of these networks are given in Appendix~\ref{app:architecture}.

The outgoing wavefunction is entirely determined by the single function $f(\theta,\phi)$, defined on the unit sphere. From this function we construct
\begin{equation} 
	\tilde \psi_{\mathrm{out}}(r,\theta,\phi) = \tilde f(\theta,\phi) \frac{e^{ikr}}{r} + \frac{\Delta_{\Omega}\tilde f(\theta,\phi)}{2ik} \frac{e^{ikr}}{r^2}\text.
\end{equation}
Here $\Delta_\Omega$ denotes the Laplace-Beltrami operator on the unit sphere. This form is imposed in part by the requirement that the loss function $\|(k^2/2M - \hat H)\tilde\psi\|_1$ be finite. If the $r^{-2}$ term were neglected, the Schr\"odinger violation $(E-\hat H)\tilde\psi$ would have behavior proportional to $r^{-3}$, and so its $L_1$ norm would be divergent. See Appendix~\ref{app:asymptoticstate} for derivations of the required asymptotic behavior as $r\rightarrow \infty$. The function $\tilde f$ is constructed as a neural network, detailed in Appendix~\ref{app:architecture}.

The wavefunction near the origin, $\tilde \psi_{\mathrm{bulk}}$, is also constructed from an MLP as detailed in Appendix~\ref{app:architecture}. The precise architecture of this wavefunction does not matter for the correctness of the algorithm as long as it decays sufficiently fast. Some architectures lead to lower $L_1$ losses, and therefore more precise estimates, than others. The architecture used in this section was chosen in part because it results in a visible discrepancy between the exact cross section and the one extracted from the neural scattering states, providing an example of the meaning of the error estimates.

The parameterization of these neural networks lifts the functional $\mathcal L[\tilde\psi]$ of wavefunctions, to a function $\mathcal L(\lambda)$ of the network parameters. We train the scattering state by performing stochastic gradient descent---specifically, \texttt{Adam}~\cite{kingma2014adam}---on $\mathcal L(\lambda)$. The loss is estimated by taking samples in $\mathbb R^3$ from a (normalized) sampling distribution $p(\cdot)$:
\begin{equation}\label{eq:l1norm_num}
	\mathcal{L}(\lambda)
	\approx \frac{1}{N_s} \sum_{\bm{r}} \frac{| (E-\hat H)\tilde\psi_\lambda(\bm{r})|}{p(\bm{r})}
	\text.
\end{equation}
The sum is taken over all $N_s$ samples from the distribution $p(\cdot)$. The choice of distribution $p$ affects the quality of these estimators but no other aspect of the algorithm. For this example we adopt an exponential distribution:
\begin{equation}
	p(x,y,z) \propto (x^2+y^2+z^2)^{-1} e^{-\frac{\sqrt{x^2 + y^2 + z^2}}{R}}\text.
\end{equation}
The remaining free parameter is selected to be $R=2$ (in units where the range of the potential is $r_0 = 1$).

We take a batch size of $N_s = 2^{10}$. To accelerate learning we adopt a linear-decay learning schedule, beginning with $10^{-2}$ and ending with a learning rate of $10^{-4}$. After training is complete, we estimate the cross section by numerically integrating the asymptotic behavior:
\begin{equation}
	\tilde\sigma = \int \sin\theta d\theta\,d\phi\,|\tilde f(\theta,\phi)|^2
	\text.
\end{equation}
On the left panel of Fig.~\ref{fig:sphere}, we show the cross section computed from the neural scattering states. Statistical errors in this estimate are negligible, and not shown. The dominant error is systematic, coming from the failure of $\tilde\psi$ to be an exact scattering state.

As discussed above, the value of the loss $\mathcal L(\lambda)$ provides an upper bound on the possible error in the asymptotic behavior of $\tilde\psi_\lambda$. To be precise, the true and approximated cross sections, $\sigma$ and $\tilde\sigma$, obey~\cite{Lawrence:2026wrf}
\begin{equation}
	|\sqrt\sigma-\sqrt{\tilde\sigma}| \le \frac{1}{2\sqrt\pi}M \|\xi\|_\infty \mathcal L[\tilde\psi]
	\text.
\end{equation}
Here $M$ is the mass of the particle being scattered, and $\xi$ is the true scattering eigenstate. The notation $\|\xi\|_\infty$ indicates the $L_\infty$ norm of the wavefunction $\xi$, i.e.~the maximum amplitude of the state.

The norm $\|\xi\|_\infty$ is difficult to determine \emph{a priori}. As a property of the true scattering state, finding this value is tantamount to solving the system (in which case neural scattering states would not be needed). However, there are several ways to approximate this value without knowing it exactly. One option is to inspect the neural scattering state itself. We take this approach to construct the error estimates shown in the left panel of Figure~\ref{fig:sphere}. After training is complete, we sample $2^{18}$ points $x_n$, evaluate $\tilde\psi(x_n)$ on each, and take the maximum magnitude as an approximation to $\|\xi\|_\infty$. This is an approximation in two ways: first, there is no guarantee that we got close to the true maximum of $|\tilde \psi|$, and second, the maximum of $|\tilde\psi|$ need not be similar to that of $|\xi|$, even when the $L_1$ loss is small. Regardless, we find that this approach performs well in practice, behaving as a moderately conservative estimate of the systematic error in the calculation.

Another approach is possible. The right-hand panel of Fig.~\ref{fig:sphere} shows, over the course of the training of $\tilde\psi$ (at an energy $E=1$), the loss $\mathcal L$ compared with the error in the cross section. In this case, the linear relation between the two is clearly visible, and the coefficient $\|\xi\|_\infty$ can be estimated from the slope.


\section{Neutron-proton phase shifts}\label{sec:np}
Now we turn to considering realistic nuclear forces. The simplest scattering process from a computational perspective is neutron-proton scattering, which is elastic and involves only distinguishable particles. Such two-body scattering is equivalent (by the reduced-mass formalism) to one-body scattering from a central potential as considered in the previous section. For this reason we will see that the formalism and computational setup need only minimal modification.

The Hamiltonian governing neutron-proton scattering has the form
\begin{equation}
	\hat H = -\frac{\hat \nabla_p^2}{2M_p}  -\frac{\hat \nabla_n^2}{2M_n} + \hat V(x_p - x_n)\text,
\end{equation}
where $x_{p,n}$ are 3-dimensional coordinates for the proton and neutron. The proton and neutron masses are respectively $M_p,M_n$. The function $V$ is matrix-valued: $V(\cdot)$ is a $4 \times 4$ matrix acting on the spins of the two nucleons. For the potential $V$, we take the Argonne $v_6'$~\cite{Wiringa:2002ja} potential, which has one-pion-exchange $v^{\pi}$ and short range potentials $v^R$. In this work we neglect electromagnetic interactions. The potential contains 6 kinds of two-body spin-isospin terms; thus the potential operator can be written
\begin{equation}
	\hat V_{\mathrm{AV6'}} = v^{\pi} + \sum_{i=p}^6 v^R_p = \sum_{(i,j)}\sum_{p=1}^6 v_p(\bm{r}_{ij}) \mathcal{O}^p_{ij} 
\end{equation}
where the six spin/isospin operators are
\begin{equation}
\mathcal{O}^p_{ij} = \left[ 1, \bm{\sigma}_i\cdot\bm{\sigma}_j, S_{ij}\right] \otimes \left[ 1, \bm{\tau}_i\cdot\bm{\tau}_j\right]\text.
\end{equation}
Here $\bm{r}_{ij}$ is the relative position of two particles (labeled as $i,j$) under consideration. $\bm{\sigma}_{i,j}$ are Pauli operators acting in the spin space of $i,j$th particles, $\bm{S} = \frac{1}{2}(\bm{\sigma}_i + \bm{\sigma}_j)$, and the tensor $S_{ij}= (\bm{\sigma}_i\cdot \bm{r}_{ij})\otimes(\bm{\sigma}_j\cdot \bm{r}_{ij})/r_{ij}^2 - 3 \bm{\sigma}_i\cdot\bm{\sigma}_j$. The Pauli operators in isospin space are denoted $\tau$.
The distance dependence $v_p(\bm{r})$ depends only on the relative position $\bm{r}$ between particles, and they contain contributions from both one-pion-exchange and short-range interactions. The coefficients in each function $v_p$ are determined by projecting down from the original 18-component potential~\cite{Wiringa:1994wb}, which is in turn determined by fits to nucleon scattering data and deuteron binding.

Due to the spin degrees of freedom, the scattering state is a function $\mathbb R^6 \rightarrow \mathbb C^4$. The state is symmetric under overall translations, and as a result the $L_1$ loss of an approximate scattering state will diverge if the integral is taken over the full $\mathbb R^6$. Instead, we integrate only over the $\mathbb R^3$ defined by having the center of mass at the origin. In terms of the relative coordinate $\bm{r}$, the scattering state is written
\begin{equation}
		\psi(\bm{r}) = e^{ikr\cos\theta}(u_p \otimes u_n)
 + \psi_{\mathrm{bulk}}(\bm{r}) + \psi_{\mathrm{out}}(\bm{r}) 
\text.
\end{equation}
Above, $u_p$ and $u_n$ are the two-component polarization vectors defining the ingoing spins of the scattering process; $\psi$ itself is a four-component vector.
The relation of lab energy $E_{\mathrm{lab}}$ (the kinetic energy of the incident neutron in the lab frame) and the relative momentum $k$ is
\begin{equation}
k^2 = \frac{M_p^2 \; E_{\mathrm{lab}} \; (E_{\mathrm{lab}}+ 2 M_n)}{(M_p+M_n)^2 + 2\;E_{\mathrm{lab}}\;M_p}\;\text.
\end{equation}

The architecture of the neural network parameterizing $\tilde \psi_{\mathrm{bulk}}$, which has eight real outputs (folded into a four-component complex vector), is detailed in Appendix~\ref{app:architecture}. We need the loss functional to be finite: the error must decay sufficiently quickly with radius. It is sufficient to ensure that the error $|(E-\hat H)\psi|$ is $O(r^{-4})$, to which end the asymptotic outgoing state is given the form
\begin{equation}
	\tilde \psi_{\mathrm{out}} =  \tilde f(\Omega,s_p,s_n) \frac{e^{ikr}}{r} + \frac{\Delta_{\Omega}\tilde f(\Omega,s_p,s_n)}{2ik} \frac{e^{ikr}}{r^2}\text.
\end{equation}
For this section, instead of parameterizing $\tilde f$ by a neural network, we write it explicitly as a sum of low-lying spherical harmonics, namely the $s$, $p$, and $d$-waves (i.e.~$\ell = 0,1,2$):
\begin{equation}
	\tilde f(\Omega) = \sum_{l=0,m}^{l_{\mathrm{max}}} \tilde f_{l,m} Y_l^m(\Omega)
	\text.
\end{equation}
The trainable coefficients are the $\tilde f_{l,m}$. In this case the angular Laplacian of $\tilde f$ can be computed analytically, and the outgoing state becomes
\begin{equation}
	\tilde \psi_{\mathrm{out}} =  \sum_{l=0,m}^{l_{\mathrm{max}}}  \tilde f_{l,m} Y_l^m(\Omega) \left[ \frac{e^{ikr}}{r} -  \frac{l(l+1)}{2ik} \frac{e^{ikr}}{r^2} \right]\;\text.
\end{equation}
The mechanistic generalization of this approach to arbitrary dimensions is outlined in Appendix~\ref{app:laplacian}.

\begin{figure}
	\includegraphics[width=0.95\linewidth]{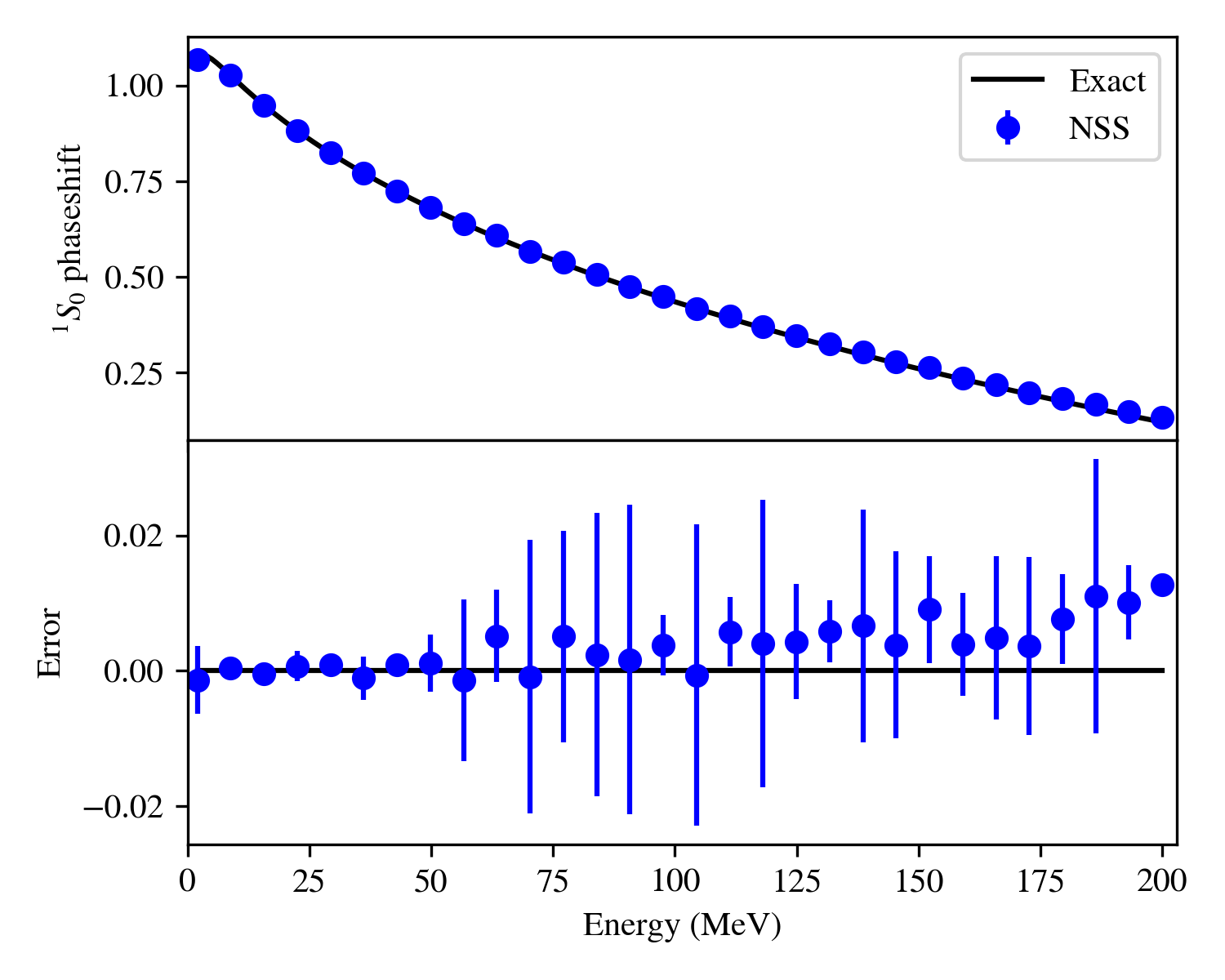}
	\caption{The ${}^{1}\!S_0$ phase shift of neutron-proton scattering, as a function of energy. Nearly exact numerical results, obtained from integrating the radial Schr\"odinger equation, are shown as the solid line. The blue data is extracted from learned neural scattering states. Systematic errors are approximated from the degree of unitarity violation, as described in the text.\label{fig:NP}}
\end{figure}

The procedure for training is the same as in the previous section. The results of this computation are shown in Figure~\ref{fig:NP}. In this case we plot the $^{1}\!S_0$ phase shift as a function of the kinetic energy in the lab frame. We obtain excellent agreement with the nearly exact results obtained by integrating the radial Schr\"odinger equation, using the same (Argonne v6') potential. 

As before, the dominant sources of error are not statistical in nature, but instead are from the choice of ansatz used. In this case, systematic error estimates obtained from the results of~\cite{Lawrence:2026wrf} are overly conservative (and therefore not shown in the figure). For this calculation we introduce another method of estimating systematic errors, exploiting the fact that neural quantum states do not, by default, respect unitarity.

In Figure~\ref{fig:NP} we have plotted a phase shift. This phase shift is not extracted directly from the NSS. From the NSS what we extract is (an approximation of) the partial-wave amplitude $e^{2 i \delta}$. Unitarity tells us that the magnitude of this complex number should be exactly unity, but the NSS does not have unitarity as a hard constraint, and so the magnitude extracted differs somewhat from $1$. To estimate the errors on the phase shift we make the following assumption: the errors in the partial-wave amplitude are Gaussianly distributed on the complex plane, with covariance proportional to the identity. This means that the error in the magnitude is typically of the same size as the error in the phase shift.

This assumption is tested in Figure~\ref{fig:NP}. The error bars plotted in that figure are given by twice the error in the magnitude of the partial-wave amplitude. For individual points we find that these error estimates are almost always on the conservative side. Moreover, looking across the set of all points, the typical size of errors in the magnitude is indeed the same as the typical size of errors in the phase shift.


\section{Neutron-deuteron cross sections}\label{sec:nd}

Nucleon-nucleon scattering is computationally trivial, as the Schr\"odinger equation may be reduced by the partial wave expansion to a set of coupled ODEs, namely the radial Schr\"odinger equation. The comparison curve in Figure~\ref{fig:NP} was orders of magnitude cheaper to produce than the NSS data points, and is more precise. The motivation for neural scattering states is the possibility that they may generalize to many-body scattering problems, which can no longer be treated by a low-dimension ODE.

The simplest nuclear scattering problem with $A > 2$ is nucleon-deuteron scattering ($A=3$). We continue to neglect Coulomb forces, and so we consider neutron-deuteron scattering where their effect is smallest. In this section we study the performance of neural scattering states in reproducing elastic and inelastic neutron-deuteron cross sections. The neutron-deuteron scattering problem has been studied via various \textit{ab initio} approaches, most notably with the Faddeev formalism~\cite{PhysRev.81.761, Kloet1973ElasticNS, WOAmrein_1982, PhysRevC.49.R14, Hber1995PhaseSA, Payne:1999je, Ishikawa:2000dq, Miller:2022beg}. These calculations have shown excellent agreement with experimental measurements. Our approach does not outperform these calculations, but opens the door to the generalizations at substantially larger $A$.

Elastic and inelastic bound-state scattering each introduce additional complications over nucleon-nucleon scattering, and so we discuss each in turn in the subsections that follow.

\subsection{Elastic scattering}\label{ssec:elastic}

As in the case of nucleon-nucleon scattering dealt with previously, the wavefunction is split into three terms: an ingoing plane wave, an outgoing spherical wave, and a ``bulk'' wavefunction to be parameterized by a neural network. The ingoing plane wave consists of a neutron and a deuteron, and in the center-of-mass frame reads
\begin{equation}\label{eq:nd-in}
	\begin{split}
		\psi_{\mathrm{in}} ={}& u_n \otimes
	e^{ik\cdot\left(x_{1p} - x_2\right)} \Psi(x_1 - x_p)\\
		&-
		\hat S \left[u_n \otimes e^{ik\cdot\left(x_{2p} - x_1\right)} \Psi(x_2 - x_p)\right]
	\text.
	\end{split}
\end{equation}
Here $x_p,x_1,x_2$ are respectively the coordinates of the proton and the two neutrons. The center of mass of the first neutron and the proton is denoted $x_{1p} = \frac{M_n x_1 + M_p x_p}{M_n + M_p}$, and $x_{2p}$ is defined similarly.
The momentum $k$ is given in terms of the lab energy by
\begin{equation}
	k = \frac{\sqrt{2 E_{\mathrm{lab}} M_n}}{\frac{M_n}{M_n + M_p} + 1}
	\text.
\end{equation}
All terms in the wavefunction have eight components due to the three dynamical fermions. There are $2_{\mathrm{neutron}} \times 3_{\mathrm{deuteron}} = 6$ possible spin states of the ingoing neutron and deuteron. The ingoing neutron spin is selected by the $2$-component spin polarization vector $u_n$, and the deuteron spin is implicit in the choice of wavefunction $\Psi$ (itself a $4$-component object). Finally, the operator $\hat S$ swaps the spins of the two neutrons, so that $\psi_{\mathrm{in}}$ is properly antisymmetric as written. This state is an eigenstate of a Hamiltonian which contains an interaction between the two constituent nucleons of the deuteron, but not between either of them and the free neutron.

Below the deuteron break-up threshold, the only outgoing channels are elastic, consisting again of a neutron and a deuteron. At large separations, this is a two-body scattering problem, and therefore the outgoing behavior of the wavefunction may be parameterized in the same manner as for nucleon-nucleon scattering. To be precise, we again write
\begin{equation}\label{eq:out-elastic}
	\tilde \psi_{\mathrm{out}}
	= \frac{e^{ik|x_r|}}{|x_r|} \tilde f_{\mathrm{nd}} [1-w(|x_r|)] \Psi(x_d)
	+\text{(antisym.)}
	\text,
\end{equation}
where $f_{\mathrm{nd}}$ depends only on angular variables, $x_r$ is the distance between the deuteron center of mass and the free neutron, and $x_d$ is deuteron internal coordinate. The window function $w(r)$ is given by
\begin{equation}\label{eq:window}
	w(r) = \frac{e^{-r/r_0}}{e^{r/r_0} + e^{-r/r_0}}
	\text.
\end{equation}
In the discussion of the numerical results below we will consider several different values of $r_0$ as a strategy for quantifying systematics. Typically we find that $r_0 \approx 5\,\mathrm{fm}$ gives a good tradeoff between efficient training and small final loss.

The bulk wavefunction is no longer a function only of one displacement, but now of two. As before we parameterize it with a multi-layer perceptron, multiplied by the same window function $w(r)$. The details of the construction of the perceptron are given in Appendix~\ref{app:architecture}. This completes the specification of the NSS ansatz, except for some difficulties related to the deuteron wavefunction $\Psi$, which we now address.

\begin{figure*}
	\includegraphics[width=0.32\linewidth]{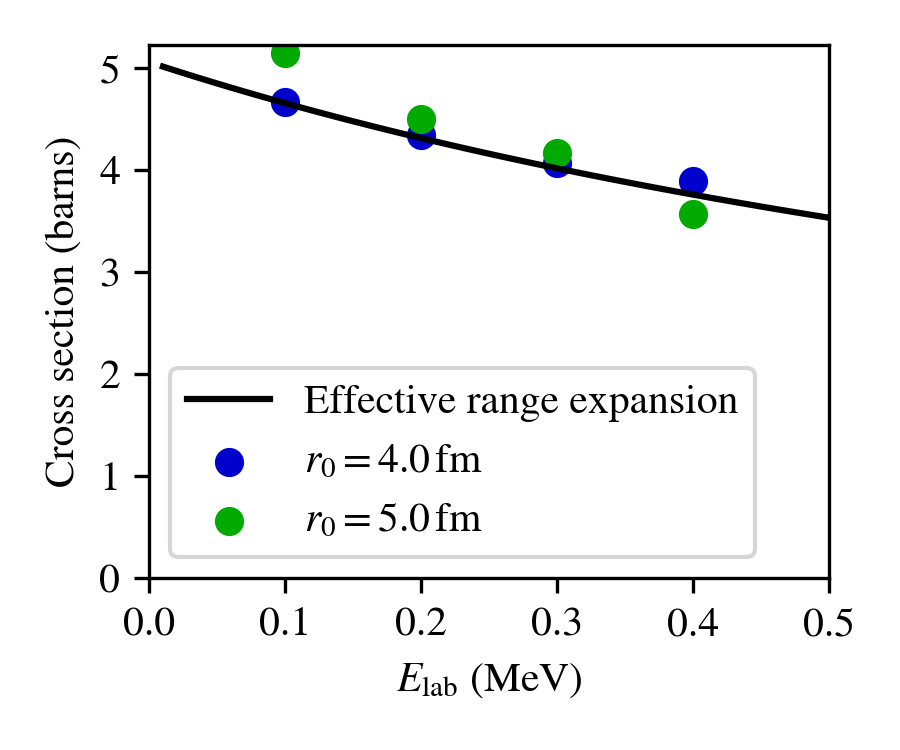}
	\hfil
	\includegraphics[width=0.32\linewidth]{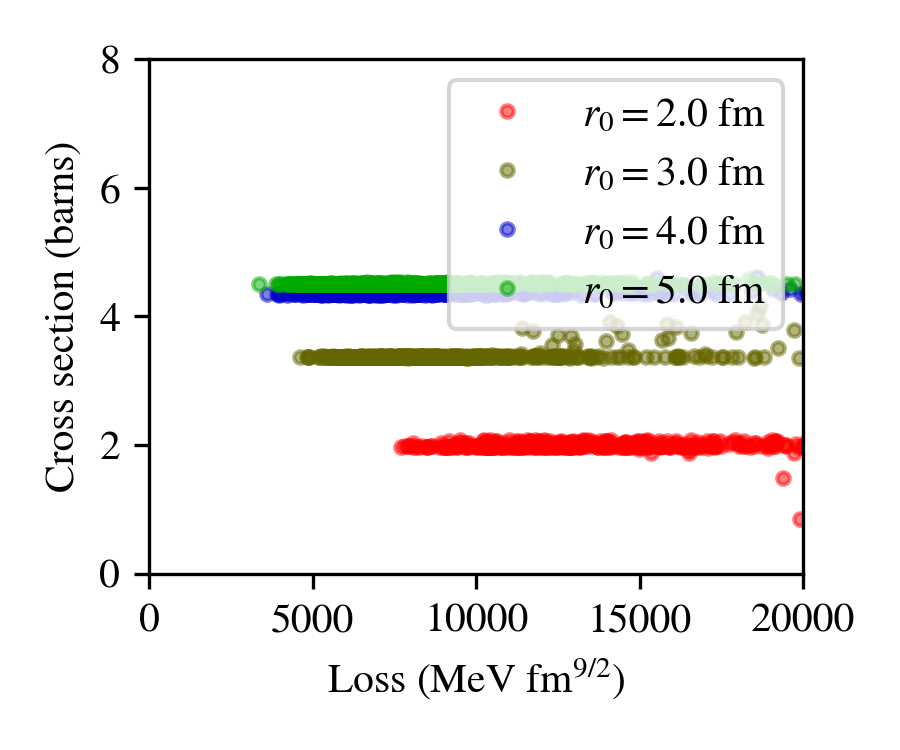}
	\hfil
	\includegraphics[width=0.32\linewidth]{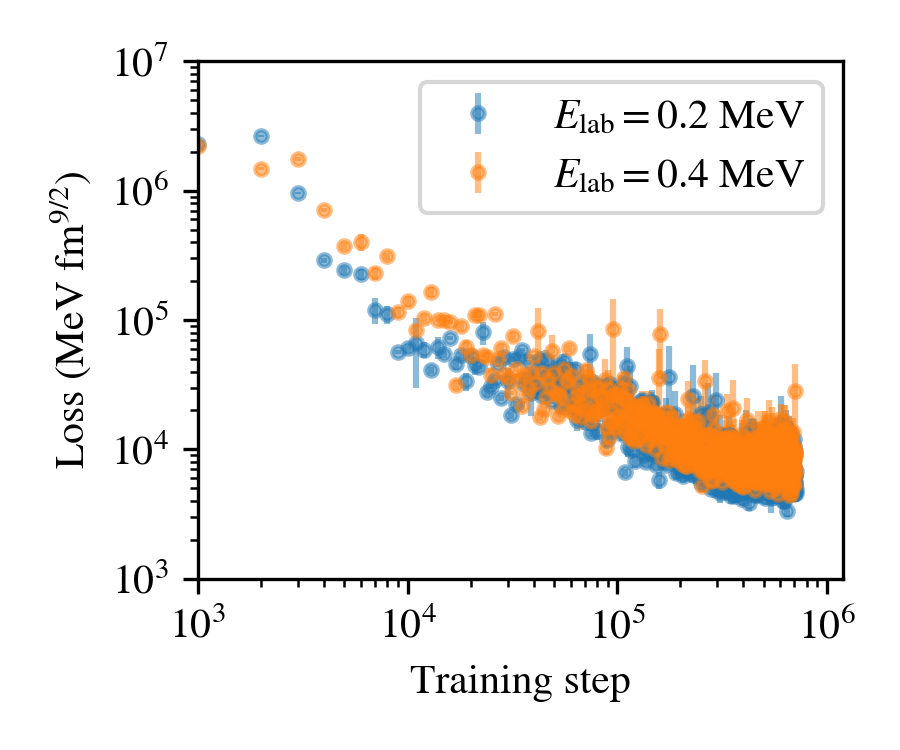}
	\caption{Calculation of the neutron-deuteron cross section in the quartet channel via neural scattering states. The NSS calculation shows good agreement with the prediction of the effective-range expansion, computed in~\cite{Witala:2003mr} from the Argonne V18 potential (left). The center plot shows the relation between the loss and the calculated cross section, at $E_{\mathrm{lab}} = 2\,\mathrm{MeV}$, for various window sizes. The training history of two runs, both with window size $r_0 = 5\,\mathrm{fm}$, is shown on the right. \label{fig:nd-elastic}}
\end{figure*}

Computationally, we do not have access to the exact deuteron wavefunction. This presents a serious problem in the estimation of the loss function. Formally, the $L_1$ norm of the Schr\"odinger violation is guaranteed to be infinite, as the error in the deuteron state introduces a (small) violation at every point in both the ingoing plane wave and the outgoing spherical wave. The algorithm used previously therefore amounts to performing stochastic gradient descent on a divergent integral\footnote{One might hope that this algorithm works anyway, on the grounds that sampling points far from the origin is uncommon. Whether or not one can ``get away with'' such an approach depends in part on the gradient descent algorithm used. Many algorithms, including \texttt{Adam}, involve some form of ``gradient clipping'' either explicitly or implicitly: occasional anomalously large gradients are treated as if they have a smaller magnitude. This has the effect of introducing a sort of regulator on the estimator, replacing the divergent integral by a similar-looking integral which converges. As a result, gradient descent using the formally divergent loss may well perform acceptably in some cases.}.

Our approach for handling the inexactness of the available bound-state wavefunctions was introduced in~\cite{Lawrence:2026wrf}. The original loss functional $\mathcal L$ defined in Eq.~(\ref{eq:L1norm_ge}) depends on the approximated scattering state $\tilde \psi$ alone. In turn, the approximated scattering state $\tilde\psi$ is constructed from an approximated bound-state wavefunction $\tilde\Psi$, bulk behavior $\tilde\psi_{\mathrm{bulk}}$, and an outgoing wave $\tilde f$. We begin by defining a new object denoted $\mathcal L_{\tilde \Psi}$, in which these dependences are made explicit:
\begin{equation}
	\mathcal L_{\tilde\Psi}[\tilde\psi_{\mathrm{bulk}},\tilde f] =
	\mathcal L[\tilde\psi(\tilde\Psi, \tilde \psi_{\mathrm{bulk}}, \tilde f)]\text.
\end{equation}

Now consider the evaluation of this loss functional in the case $\tilde\Psi = \Psi$---that is, look specifically at the two-argument functional $\mathcal L_{\Psi}$. This is the functional we would have computed if the exact bound-state wavefunction were available. The minimum of $\mathcal L_\Psi[\cdot,\tilde f] = 0$ is obtained where $\tilde f$ is the true outgoing wave. Unfortunately without access to $\Psi$ we cannot evaluate $\mathcal L_\Psi$.

We can, however, approximate $\mathcal L_\Psi$ with bounded error. Intuitively, the idea is to drop, from $\mathcal L_{\tilde\Psi}$, a set of terms which ``should'' vanish, and only fail to vanish because of the inexactness of $\tilde\Psi$. Concretely, the approximated loss is defined to be:
\begin{widetext}
\begin{equation}\label{eq:loss-approx-bound}
	\tilde {\mathcal L}[\tilde\psi_{\mathrm{bulk}}, \tilde f] :=
	\int
	\left|
	-\hat V_{\mathrm{ext}} \psi_{\mathrm{in}}
	+
	(E - \hat H) \tilde\psi_{\mathrm{bulk}}
	+
	(E - \hat H) \tilde\psi_{\mathrm{out}}
	- (E_{\mathrm{deut}} - \hat H_{\mathrm{deut}})\tilde\psi_{\mathrm{out}}
	\right|
	\text.
\end{equation}
\end{widetext}
Here $E_{\mathrm{deut}}$ is the (negative) binding energy of the deuteron, and $\hat H_{\mathrm{deut}}$ consists of all those terms in the Hamiltonian which concern the free deuteron: the relative momentum, and the interaction between the neutron and the proton. Two of the remaining terms---the interactions between the free neutron and the two particles in the deuteron---are captured by the potential $\hat V_{\mathrm{ext}}$.

Note that this loss function can no longer be written in terms of $\tilde \psi$ itself. We are no longer considering a minimum principle over the space of scattering states $\tilde\psi$. Instead we have a minimum principle over tuples $(\tilde\psi_{\mathrm{bulk}},\tilde f)$.

As shown in~\cite{Lawrence:2026wrf}, this loss function has the property that it is a bounded approximation of $\mathcal L_\Psi$. In particular, there is some (potential-dependent) constant $C_{\tilde\Psi}$ such that, for any $\tilde\psi_{\mathrm{bulk}}$ and $\tilde f$,
\begin{equation}
	\left|\tilde {\mathcal L}[\tilde\psi_{\mathrm{bulk}},\tilde f]
	- \mathcal L_{\tilde\Psi}[\tilde\psi_{\mathrm{bulk}},\tilde f]
	\right|
	\le C_{\tilde\Psi}
	\text.
\end{equation}
For the purposes of this work, we adopt a deuteron wavefunction which is close enough to the true wavefunction that this error is a negligible contribution to the loss function, and we do not attempt to estimate $C_{\tilde\Psi}$ itself. Nevertheless we have found, particularly for less-accurate deuteron wavefunctions, that the convergence of the algorithm relies on the approximated loss $\tilde {\mathcal L}$ being used in the gradient descent.

To demonstrate elastic bound-state scattering, we consider the Argonne v6' potential as in the previous section, and search for an approximate scattering state representing neutron-deuteron scattering.
The scattering cross section is extracted directly from the asymptotic function $\tilde f$, which depends only on a single coordinate giving the separation between the neutron and deuteron. The (approximated) cross section is given by an integral over the celestial sphere in this coordinate:
\begin{equation}
	\tilde\sigma
	= \int_{S^2} |\tilde f|^2
	\text.
\end{equation}

Partly for technical reasons discussed in the next subsection, we restrict outgoing behavior to the $s$-wave alone; that is, $\tilde f$ is a constant. This is an excellent approximation at energies below $1$ MeV. We compute the polarized cross section, for a neutron in the $+\frac 1 2$ state and a deuteron in the $+1$ state. This corresponds to the quartet channel of spinor-vector scattering, and is the dominant contribution to unpolarized scattering at low energies.

Figure~\ref{fig:nd-elastic} shows the results of the NSS calculation at these low energies. At left the cross section extracted from the NSS is compared to the prediction, for the quartet channel, from the effective range expansion, as reported in~\cite{Witala:2003mr}. Each calculation is performed for window sizes of $r_0 = 4.0\,\mathrm{fm}$ and $r_0 = 5.0\,\mathrm{fm}$. We find good agreement across this range of energies, where the effective range expansion and the restriction to $s$-wave asymptotics are both good approximations. Moreover, the spread between calculated cross sections coming from different ansatz parameters gives a good estimation of the systematic errors.

In the center panel, the loss is plotted against the estimated cross section, for several different training runs. All training runs in this plot are at $E_{\mathrm{lab}} = 0.2\,\mathrm{MeV}$. There is one data point for every $10^3$ training steps. In all of these calculations, we find that the bound relating the error in the cross section to the $L_1$ norm of the Schr\"odinger violation is not close to being tight. In other words, the neural scattering state is considerably more accurate than would be expected from the loss alone.

Finally, the training history for two runs is shown in the rightmost panel of Figure~\ref{fig:nd-elastic}. To improve convergence we use an exponential learning rate decay. The learning rate begins as $10^{-3}$ at the first training step, and decays to $10^{-5}$ at training step $2 \times 10^5$. For all these calculations we use a batch size of $2^{12}$. The loss is evaluated by reweighted Monte-Carlo, sampling from an isotropic Gaussian with standard deviation $\frac 3 2 r_0$.

In the case of the doublet channel (which may be extracted by simulating any other spin combination), we found that the loss remained somewhat larger than the losses obtained above, and moreover that the cross section did not converge quickly to the result expected from the effective range expansion. Either more training or an improved ansatz, or both, is therefore needed to accurately reproduce the full structure of elastic neutron-deuteron scattering.

\subsection{Inelastic scattering}\label{ssec:inelastic}

In the case of inelastic scattering there are now additional outgoing channels. The elastic channels are unchanged from above and need not be discussed further. The inelastic channels may be characterized by an asymptotic function $f_{\mathrm{inel}}$. As in the two-body case this function is complex valued and defined on the celestial sphere, but the correponding celestial sphere is now that of $\mathbb R^6$---that is to say, $S^5$.

It is convenient to express the outgoing inelastic behavior in terms of Jacobi coordinates. Define $y_d$ and $y_r$ according to
\begin{equation}
	\begin{split}
		y_d &= x_1 - x_p \text{ and}\\
		y_r &= \frac{M_p + M_n}{\sqrt{M_p (M_p + 2 M_n)}} \left(x_2 - \frac{M_p x_p + M_n x_1}{M_p + M_n}\right)\text.
	\end{split}
\end{equation}
These coordinates have the property that the kinetic term (without center-of-mass motion) may be written
\begin{equation}
	\hat H_{\mathrm{kin}} = -\frac{1}{2\mu} \left(\frac{\partial^2}{\partial y_r^2} + \frac{\partial^2}{\partial y_d^2}\right)
	\text{ with }
	\mu = \frac{M_p M_n}{M_p + M_n}
	\text.
\end{equation}
One virtue of these coordinates is that the outgoing behavior can again be written as a (hyper)spherical wave with a simple form, for any short-range interaction:
\begin{equation}
	\psi_{\mathrm{out,inel}} = e^{ikr}{r^{-5/2}} f_{\mathrm{inel}}\text.
\end{equation}
Here (and in what follows) $r$ is the hyperspherical radial coordinate. The inelastic scattering data is represented by the function $f_{\mathrm{inel}}$, which is an $8$-component complex-valued function of the hypersphere $S^5$.

Some care is needed to correctly treat the asymptotic behavior in the inelastic channels. In the elastic case above we restricted to the $s$-wave (an excellent approximation at low energies). In the inelastic case this is no longer sensible: the hyperspherical $s$-wave is not a good approximation. Therefore we must now put some effort in to constructing an ansatz and loss function with reasonable behavior.

Consider an ansatz constructed only from the leading-order (in $r^{-1}$) behavior, so that
\begin{equation}
	\tilde\psi_{\mathrm{LO}} = \psi_{\mathrm{in}} + e^{ikr}{r^{-5/2}} \tilde f_{\mathrm{inel}} + O(e^{-r})\text.
\end{equation}
Far from the origin, the Schr\"odinger violation scales as $r^{-9/2}$, and therefore the integral which defines the $L_1$ loss is divergent.

The same phenomenon also happens in the elastic case, although the divergence there is only logarithmic. The leading-order term, proportional to $r^{-1}$, results in a violation proportional to $r^{-3}$. Above, we implicitly dealt with this problem by considering only the $s$-wave: all other partial wave amplitudes were set to $0$. In the elastic channel the violation is proportional to the Laplacian of $f$ (see Appendix~\ref{app:asymptoticstate}), which vanishes in the $s$-wave. Had we included any higher-order partial waves we would have found a numerically divergent loss; by restricting to the $s$-wave the loss was guaranteed to converge. The price paid is that without including higher partial waves, the loss could never be made arbitrarily close to zero no matter how high-quality the ansatz for $\tilde\psi_{\mathrm{bulk}}$.

In order for the $L_1$ norm of the violation to be finite, we need to include sufficient terms in $\tilde\psi_{\mathrm{out,inel}}$ so that the violation decays at least as $r^{-13/2}$. This is remarkably difficult to accomplish in practice, and it is worth briefly considering several alternative approaches in turn. Suppose first that we add extra terms parameterized by separate neural networks:
\begin{equation}
	\tilde\psi_{\mathrm{out},\mathrm{NNLO}} = \frac{e^{ikr}}{r^{5/2}}\left[\tilde f^{(0)}_{\mathrm{inel}} + r^{-1}\tilde f^{(1)}_{\mathrm{inel}} + r^{-2}\tilde f^{(2)}_{\mathrm{inel}}\right]\text.
\end{equation}
This is sufficient in principle---when $\tilde f^{\bullet}$ is correct, the violation scales as $r^{-13/2}$ as desired. However, when $\tilde f^{(0)}_{\mathrm{inel}}$, $\tilde f^{(1)}_{\mathrm{inel}}$, and $\tilde f^{(2)}_{\mathrm{inel}}$ do not obey precisely the correct relations, as described in Appendix~\ref{app:asymptoticstate}, the $L_1$ loss remains divergent. Numerically, exact training is not achievable, and we do not have a well-defined loss.

Suppose instead that we construct $\tilde f^{(1)}_{\mathrm{inel}}$ from $\nabla^2 \tilde f^{(0)}_{\mathrm{inel}}$, by automatic differentiation. This is algebraically the cleanest approach, but algorithmically impractical because the evaluation of $\tilde\psi$ requires fourth-order derivatives. (Therefore the evaluation of $\hat H\tilde\psi$ requires sixth-order derivatives, and its optimization requires seventh-order derivatives.)

Another approach is to abandon the parameterization of the outgoing wave by a neural network, and instead decompose explicitly as a sum of partial waves. A simple algorithm for accomplishing this---particularly the rapid evaluation of arbitrarily many Laplacians on such a decomposition---is sketched in Appendix~\ref{app:laplacian}. Unfortunately, inelastic scattering states do not have rapidly convergent hyperspherical partial wave expansions, so a good calculation should be expected to require impractically many terms in this expansion. 

Instead, we generalize the approach taken in the elastic case to the problem of
an inexact bound-state wavefunction. There, we dropped from the loss function
terms which, while infinite, appeared only because we had an approximate
bound-state wavefunction. In the inelastic case we will do the same, dropping
terms related to the subleading (in $r$) behavior of the outgoing wave. In both
cases, we begin by constructing an ideal loss function, which is finite, but
beyond our ability to compute. We then show that this ideal loss function may be approximated, with bounded error, by a different loss function which we are able to compute.

Denote the leading-order outgoing amplitude $\tilde f^{(0)}$. In principle, the
free Schr\"odinger equation determines functions $\tilde f^{(j)}$ for all $j \ge 1$ (see Appendix~\ref{app:asymptoticstate}) such that the Schr\"odinger violation of the series
\begin{equation}
	\tilde \psi_{\mathrm{out}} = \bar w(r) e^{ikr} r^{\frac{1-d}{2}} \sum_{j=0}^\infty r^{-j} f^{(j)}
\end{equation}
vanishes to all orders. Above we have denoted $1-w = \bar w$ for brevity. (The window function defining the bulk is $w$, and $\bar w$ is used for the outgoing behavior.) Henceforth we take this property to be the definition of $\tilde f^{(j)}$, and furthermore we write
\begin{equation}
	\tilde F =  r^{\frac{1-d}{2}} \sum_{j=0}^\infty r^{-j} \tilde f^{(j)}.
\end{equation}
Alternatively one may consider $\tilde F$ to be defined by its limit at large $r$ and a differential equation:
\begin{equation}\label{eq:tildeF-defining}
	\lim_{r\rightarrow\infty} r^{\frac{d-1}{2}} \tilde F = f^{(0)} \quad \text{and}\quad 0 = (k^2 + \nabla^2)\left(e^{ikr} \tilde F\right)
	\text.
\end{equation}
Implicitly, $\tilde F$ is a functional of $\tilde f^{(0)}$, and we may use it to construct a loss functional $\mathcal L[\tilde \psi_{\mathrm{bulk}},\tilde f^{(0)}]$. This construction is analogous to the one in the previous section:
\begin{align}\label{eq:loss-approx-outgoing}
	\mathcal L[\tilde \psi_{\mathrm{bulk}}, \tilde f^{(0)}]
		&=
	\int \left| \Lambda \right|\\
		\text{where } \Lambda &= 
	(E - \hat H) \left(\psi_{\mathrm{in}} + \tilde\psi_{\mathrm{bulk}} + \bar w(r) e^{ikr} \tilde F\right)
		\text.\nonumber
\end{align}
For the practical reasons discussed above, $\tilde F$ is computationally inaccessible, and therefore so is $\mathcal L$. Instead we have access to some function $\tilde G$, also a power series in $r^{-1}$:
\begin{equation}
	\tilde G =  r^{\frac{1-d}{2}} \sum_{j=0}^\infty r^{-j} \tilde g^{(j)}
	\text.
\end{equation}
The only guaranteed relation between the two series is that, by definition, $\tilde g^{(0)} = \tilde f^{(0)}$.

Now consider the Schr\"odinger violation $\Lambda$ of $e^{ikr} \bar w \tilde F$. Working in spherical coordinates in arbitrary dimension $d$, we evaluate
\begin{equation}
	\begin{split}
	(k^2 + \Delta) (e^{ikr} \bar w \tilde F)
		={}& e^{ikr} \Big[(d-1) r^{-1} \tilde F \partial_r \bar w\\ + 2 \partial_r \bar w i k \tilde F &+ 2 \partial_r \bar w \partial_r \tilde F + \tilde F \partial_r^2 w\Big]
	\text.
	\end{split}
\end{equation}
The terms not involving at least one derivative of the window function vanished due to Eq.~(\ref{eq:tildeF-defining}). We now replace $\tilde F \rightarrow \tilde G$ in the above expression to obtain an approximated violation $\tilde \Lambda$:
\begin{widetext}
\begin{equation}
	\tilde \Lambda
	= (E - \hat H) (\psi_{\mathrm{in}} + \tilde \psi_{\mathrm{bulk}}) + \frac{e^{ikr}}{2M} \left[(d-1) r^{-1} \tilde G \partial_r \bar w + 2 \partial_r \bar w i k \tilde G + 2 \partial_r \bar w \partial_r \tilde G + \tilde G \partial_r^2 w\right] - V e^{ikr} \bar w(r) \tilde G
	\text.
\end{equation}
\end{widetext}
This in turn defines an approximated loss $\tilde {\mathcal L} = \int |\tilde\Lambda|$. This integral is convergent.

This approximated loss gives us considerable latitude to select $\tilde G$ in the design of a numerical algorithm. Any construction of $\tilde G$ will yield a convergent loss integral, and so we are free to construct $\tilde G$ in any manner computationally convenient, so long as the resulting error $|\mathcal L - \tilde{\mathcal L}|$ is tolerable. In practice we find that the choice
\begin{equation}
	\tilde G_{\mathrm{cheap}} = r^{\frac{1-d}{2}} \tilde f^{(0)}
\end{equation}
yields reasonable results with minimal computational cost, and so this choice is made for all numerical results that follow.

Other choices are possible. It is desirable if the error in the approximation to the loss can be driven to zero in some limit, without excessive computational expense. This was the case in the elastic scattering subsection above: improving the bound-state ansatz $\tilde\Psi$ drove this error to zero, as the error was proportional to $|\tilde\Psi - \Psi|$. A similar, systematic improvement of the approximated loss is detailed in Appendix~\ref{app:outgoing}. In short, one fixes an integer $m$ and parameterizes $\tilde g^{(1)},\ldots,\tilde g^{(m)}$ by neural networks. These functions are trained according to a loss function constructed from Eq.~(\ref{eq:f-recursion}). As long as $m$ is sufficiently large, the error in the approximated loss is driven to zero in the dual limit where the training of $\tilde g^{(j \le m)}$ is performed well, and the radial parameter $r_0$ in the window function $w(r)$ is taken to be large.

To demonstrate the calculation of inelastic scattering via neural scattering states we use the same system as before, but at higher energies so that the inelastic channel is opened. The full scattering ansatz we use is
\begin{equation}\label{eq:inelastic-ansatz}
	\tilde\psi = \psi_{\mathrm{in}} + \tilde\psi_{\mathrm{bulk}} + \tilde\psi_{\mathrm{out,el}}
	+ e^{ikr} r^{-5/2} \tilde f_{\mathrm{inel}}
	\text.
\end{equation}
The first three terms are identical to those used for the elastic calculation, and the computation of the elastic cross section is not modified from the previous subsection.
In order to obtain the inelastic cross section, we note that when the kinetic term in the Hamiltonian is written $\frac 1 2 M^{-1}_{ij} \nabla_i \nabla_j$, the conserved current of the system is
\begin{equation}
	j_i = M^{-1}_{ij} \Im \psi^\dagger \nabla_j \psi\text.
\end{equation}
The inelastic cross section is given by the ratio of an integral of this current on the celestial hypersphere, to the value of this current in the ingoing plane wave. Defining $j^{(\mathrm{in,nd})}$ to be the neutron-deuteron current in the ingoing wave and $\tilde j^{(\mathrm{inel})}$ the current constructed from the outgoing inelastic channel, we compute the cross section as
\begin{equation}
	\tilde\sigma
	= \frac{\int_{S^5} \hat n \cdot \tilde j^{(\mathrm{inel})}}{j^{(\mathrm{in,nd})}}
	\text.
\end{equation}
The integral in the numerator is taken over the celestial sphere. The inelastic-channel current may be computed directly from the final term of the scattering ansatz. Denoting that term $\tilde\psi_{\mathrm{out,inel}}$, this current is
\begin{equation}
	\tilde j^{(\mathrm{inel})} =
	M^{-1}_{ij} \Im \tilde\psi_{\mathrm{out,inel}}^\dagger \nabla_j \psi_{\mathrm{out,inel}}
	\text.
\end{equation}
The neutron-deuteron current in the denominator requires integrating over the internal deuteron coordinate. In terms of the Jacobi coordinates this is an integral over $y_d$:
\begin{equation}
	j^{(\mathrm{in,nd})} =
	\int dy_d\,j^{(in)}
	= \sqrt{\frac{2 E_{\mathrm{lab}}}{M_n}}
	\text.
\end{equation}
Jacobi coordinates have the added convenience that the mass matrix $M$ is made diagonal; however the above expressions for currents and the cross section are coordinate-independent.

\begin{figure}
	\includegraphics[width=0.95\linewidth]{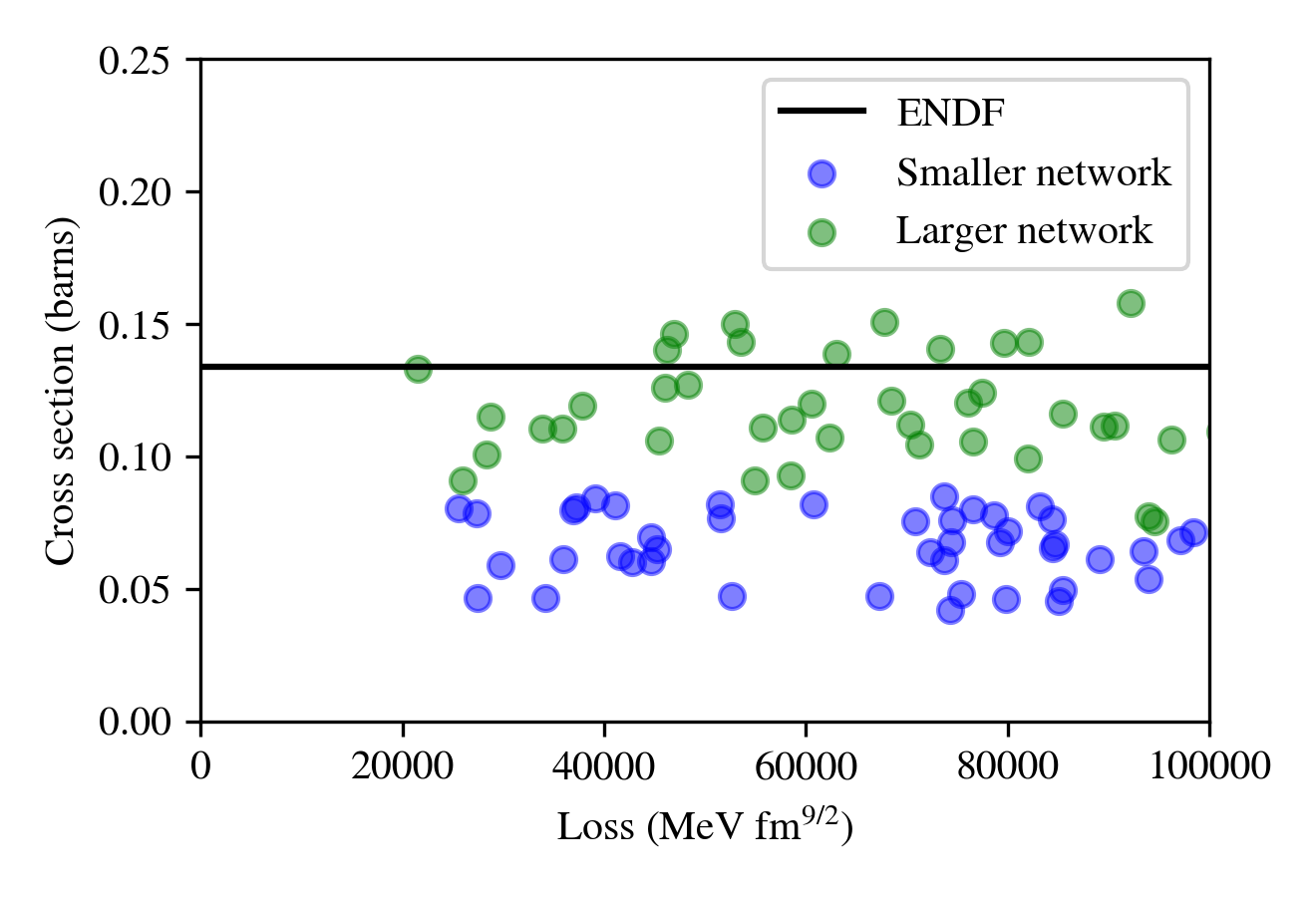}
	\caption{Calculation of neutron-deuteron scattering at $10\,\mathrm{MeV}$ incident neutron energy. The solid line, given for a loose comparison, shows the unpolarized inelastic cross section extracted from ENDF/B-VII.1~\cite{NNDC_Sigma_ENDF}. The NSS calculations are performed for polarized scattering as described in the text.\label{fig:nd-inelastic}}
\end{figure}

There is one additional source of divergence in the $L_1$ norm of the Schr\"odinger violation. Where two outgoing particles are close to each other, the interaction contributes a term to the violation which does not decay with the radial coordinate $r$. We do not handle this divergence, which is closely related to final-state interaction. Instead we run gradient descent directly on the estimator of the (formally divergent) $L_1$ integral. A more systematic approach to treating final-state interactions in neural scattering states is left for future work.

With this final approximation made, the results of an NSS calculation of neutron-deuteron scattering at $10\,\mathrm{MeV}$---well above breakup energy---are shown in Figure~\ref{fig:nd-inelastic}. We compute the polarized inelastic cross-section, where the deuteron is in the spin $+1$ state and the neutron in the $+\frac 12$ state. We use a window size of $r_0 = 2\,\mathrm{fm}$ (far too small to yield high-precision results~\cite{Glockle:1992zz}). Two different sizes of bulk MLP parameterization are used: the smaller has a depth of $2$ and a width of $250$ (as described in Appendix~\ref{app:architecture}), and the larger has a depth of $3$ and a width of $400$. Each is trained for $6.8 \times 10^4$ steps with a batch size of $2^{12}$. Again the loss is evaluated by reweighted Monte Carlo, sampling from a Gaussian of standard deviation $\frac 3 2 r_0$.

We find that the inelastic cross section is reliably reproduced to within a factor of two, and there are indications of systematic improvement as larger networks are trained. The elastic cross section---not shown in the figure---performs less well, being found far below its experimental value (around $30\,\mathrm{fm^2}$ when it should be nearly three times as large), and with less improvement as the size of the bulk network is increased. The inelastic cross section of the other two non-symmetry-related polarizations---$(0,\frac 1 2)$ and $(-1,\frac 1 2)$---are not shown. We find that the cross section for $(-1,+\frac 1 2)$ is statistically indistinguishable, whereas the cross section for $(0,+\frac 1 2)$ is smaller by a factor of two.

This poor performance is reflected in part in the large value of the loss. Comparing to Figure~\ref{fig:nd-elastic}, the best loss achieved in the inelastic calculation is a factor of five larger than the best loss achieved in the elastic calculation.


\section{Discussion}\label{sec:discussion}

We have presented a machine-learning method, based on neural quantum states, for determining few-body scattering properties, including cross sections and phase shifts. The method is able to model both elastic and inelastic scattering, and we have demonstrated it on both nuclear two- and three-body problems. Crucially, the method is \emph{time-independent}, meaning that approximating a long time evolution is not necessary.

By minimizing the $L_1$ norm of the Schr\"odinger violation, we are able to train an approximate scattering state, from which the cross section (or partial wave amplitude, as desired) can be calculated. Our prior formal work~\cite{Lawrence:2026wrf} both establishes that the error in the cross section will necessarily fall as the $L_1$ norm decreases, and provides a bound relating the possible cross section error to the $L_1$ norm. This result gives us a reliable way to estimate (conservatively) the systematic errors in these calculations.

For cases where the error estimates provided by~\cite{Lawrence:2026wrf} are too conservative, we have introduced several other approaches to estimating systematic errors in NSS calculations. Where an exactly conserved quantity is known, the violation of this conservation law can serve as a measure of the error in the calculation. In scattering processes the total probability current is always exactly conserved, and so the amount of violation of unitarity provides a practical estimate of systematic errors. Alternatively, in Section~\ref{sec:nd} we showed that it is feasible to simply perform multiple runs with different parameters, and use the spread in the calculation to motivate error bands.

This paper introduced several technical steps that we found to be necessary in order to construct a scattering ansatz, with a finite loss function, which can be relied upon to converge (at least in the limit of infinite training) to the true scattering state. In the main text we introduced these pieces gradually, where necessary to obtain the next result; here we list them all together:
\begin{itemize}
	\item To perform scattering calculations with neural quantum states, without needing large time-evolution, we search for the time-independent scattering state (section~\ref{sec:nss}).
	\item To ensure convergence to the true scattering data as the loss function is reduced, we use the stability theorems of~\cite{Lawrence:2026wrf}, combined with the specific choice of the $L_1$ norm $\|(E-\hat H)\tilde\psi\|_1$ as our loss function.
	\item To accommodate a non-exact bound state, we define an approximation to the ideal loss function which can be computed efficiently without needing to know the true bound-state wavefunction (section~\ref{ssec:elastic}). This approximation itself has bounded error (and therefore ``composes'' well with the stability theorems underlying the overall method).
	\item Finally, to avoid the expense of high-order derivatives, we show in Section~\ref{ssec:inelastic} that one may similarly work with approximated outgoing asymptotics, again replacing the ideal loss function by an efficiently computable approximation with bounded error.
\end{itemize}

The time-independent nature of the NSS method stands in contrast to common proposals for scattering simulations on quantum computers, including for nuclear systems~\cite{Weiss:2024mie,Stetcu:2025mjw}. It is likely that, just as quantum computers can be used for the study of ground-state properties by variational methods, the minimum principle used in this work can be used to construct a quantum algorithm for studying scattering, without the requirement of a deep circuit for accurate time-evolution.

As described in Section~\ref{ssec:inelastic}, final-state interactions result in a divergent $L_1$ loss. Future work must address this properly in order to compute inelastic scattering processes with precision and trustworthy error estimates.

Throughout this work we have ignored Coulomb interactions entirely, although the basic formalism for error estimates in the Coulomb case is described in~\cite{Lawrence:2026wrf}. This is an excellent approximation for the processes we have considered in this work. At lower energies and when both particles are charged Coulomb interactions are not negligible, and future work must take them into account. Finally, just as nothing in our formalism prevents the study of Coulomb interactions, nothing prevents the simulation of three-body (or four-body, or higher) interactions.

\acknowledgments
For their comments on computational approaches to scattering, we are grateful to Tanmoy Bhattacharya, Joe Carlson, Tom Cohen, Duff Neill, Ingo Tews, and Rahul Somasundaram. S.L.~is indebted to Brian McPeak for many discussions on neural network scaling and architectures. Tom Cohen provided valuable feedback on an earlier draft of this manuscript.

S.L.~is supported by a Richard P.~Feynman fellowship from the LANL LDRD program. Y.Y.~was initially supported by the INT's U.S. Department of Energy grant No.~DE-FG02-00ER41132, and is subsequently supported by a Darleane C. Hoffman fellowship from the LANL LDRD program. Los Alamos National Laboratory is operated by Triad National Security, LLC, for the National Nuclear Security Administration of U.S. Department of Energy (Contract No.~89233218CNA000001).

This research used resources provided by the Darwin testbed at Los Alamos National Laboratory (LANL) which is funded by the Computational Systems and Software Environments subprogram of LANL's Advanced Simulation and Computing program (NNSA/DOE).


\appendix

\section{Asymptotics of scattering states}\label{app:asymptoticstate}
Assuming a potential which decays exponentially far from the origin, the asymptotic behavior of the outgoing wave of a scattering state may be decomposed in a sum of algebraically decaying terms:
\begin{equation}\label{eq:outgoing-sum}
	\psi_{\mathrm{out}} = \sum_n f_n(\Omega) e^{ikr} r^{-\gamma_n}
	\text.
\end{equation}
Here and for the rest of this appendix, the ingoing term (a plane-wave $e^{ikx}$) and all bulk terms, defined as those which decay exponentially in $r$, have no role to play and are entirely neglected.

The exponents $\gamma_n$ and the angle-dependent coefficients $f_n$ may be determined in a physically transparent way by noting that these spherical outgoing waves must be decomposable as sums of plane waves, as plane waves form a complete basis for eigenstates of the free Hamiltonian. Therefore one expects to be able to write
\begin{equation}
	\psi_{\mathrm{out}}(x)
	= \int d\Omega_k e^{i(k\cdot x)} g(k)
	\text.
\end{equation}
Here the integration is taken over the $(d-1)$-sphere of momenta $k$ which satisfy energy conservation, and the function $g(\cdot)$ is defined only on that sphere. Where the point $x$ is far from the origin, this integral may be evaluated by the stationary-phase approximation, and one finds that $g$ determines the asymptotic angular functions $f$. The exponents $\gamma_n$ are determined exclusively by the spatial dimension of the system.

While physically transparent, the stationary-phase integration can be algebraically difficult, and so it is often more convenient to solve the free Schr\"odinger equation order-by-order in $r^{-1}$. In general spatial dimension $d$, the free eigenvalue equation reads (in appropriate coordinates)
\begin{equation}
	0 = k^2 \psi + r^{-2} \Delta_\Omega \psi + r^{1-d} \partial_r \left(r^{d-1} \partial_r \psi\right)
	\text.
\end{equation}
Here $\Delta_\Omega$ denotes the Laplace-Beltrami operator on the sphere, discussed more in the following appendix. We may now use this equation to derive a recursion relation on the $\gamma_n,f_n$. The radial Laplacian, acting on each term of Eq.~(\ref{eq:outgoing-sum}), yields
\begin{widetext}
\begin{equation}
	r^{1-d} \partial_r \left(r^{d-1} \partial_r e^{ikr} r^{-\gamma} \right)
	=
	e^{ikr} r^{-\gamma} \left[
		-k^2 e^{ikr}
		+ ik (d-1-2\gamma) r^{-1}
		- \gamma (d-2-\gamma) r^{-2}
		\right]
	\text.
\end{equation}
Therefore the free Schr\"odinger equation, acting on the outgoing state, reads
\begin{equation}
	0 = e^{ikr} \sum_n r^{-\gamma_n} \left[r^{-2} \Delta_\Omega f_n
	+ ik (d-1-2\gamma_n)r^{-1} f_n
	- \gamma_n (d-2-\gamma_n) r^{-2} f_n
	\right]
	\text.
\end{equation}
	It is clear that all $\gamma_n$ will be separated only by integer values, and therefore we define $\gamma_n = \gamma - n$. We will obtain a recursion relation, with base case defined by the leading-order power $\gamma = \frac{d-1}{2}$ which appears in the outgoing wave.

	The case $(\gamma,d)=(1,3)$ is clearly special (along with $(\gamma,d) = (0,1)$). Here, for constant $f$, all terms above vanish and one obtains a spherically symmetric outgoing wave. For other values of $d$, there are subleading terms even in the $s$-wave.

The above equation may be solved order-by-order in $r$. Reorganizing accordingly, we obtain the recursion relation
\begin{equation}\label{eq:f-recursion}
	2ik(n+1) f_{n+1} = \Delta_\Omega f_{n}
	- \left(\frac{d-1}{2} + n\right)\left(\frac {d-1} 2 - n - 1\right) f_{n}
	\text.
\end{equation}
We have defined $f_n = 0$ for all $n < 0$, giving this recursion relation a unique solution once $f_0$ is specified. The two cases of interest in this work are $d=3$ and $d=6$. In the three-dimensional case the recursion reads
	\begin{equation}
		2ik(n+1) f_{n+1} = \Delta_\Omega f_n
		+ 2 ik (n - 1) f_{n-1}
		+ n (1+n) f_{n-2}
		\text,
	\end{equation}
	and is solved at NLO by
\end{widetext}
	\begin{equation}
		f_1 = \frac{1}{2ik} \Delta_\Omega f_0
		\text.
	\end{equation}
	In the six-dimensional case the recursion reads
	\begin{equation}
		2ik(n+1) f_{n+1} =
		\Delta_\Omega f_n
		- \left(\frac 5 2 + n\right)
		\left(
		\frac 3 2 - n
		\right) f_n
	\end{equation}
	and is solved at N$^2$LO by
	\begin{equation}
		\begin{split}
			f_1 &= \frac{1}{2ik} \left[\Delta_\Omega f_0 - \frac{15}{4} f_0\right]
			\text{ and}\\
			f_2 &= \frac{-1}{8 k^2} \left[
			\Delta_\Omega^2 f_0 - \frac{11}{2} \Delta_\Omega f_0
			+ \frac{105}{16} f_0
			\right]
		\text.
		\end{split}
	\end{equation}

\section{Network architectures}\label{app:architecture}

This appendix is to give full details of the architectures of the neural networks used to represent the ``near'' and ``far'' wavefunctions, respectively denoted $\psi_{\mathrm{bulk}}$ and $\psi_{\mathrm{out}}$. As a reminder, the full wavefunction of a scattering state is constructed as
\begin{equation}
	\tilde \psi(x)
	= e^{i k x}
	+ \tilde \psi_{\mathrm{bulk}}
	+ \tilde \psi_{\mathrm{out}}
\text.
\end{equation}
The first term contains the only ingoing wave, and is responsible for uniquely specifying the scattering state. The final term contains the $S$-matrix data, and (depending on the loss function used) sufficient subleading (in $r$) behavior to render the $L_1$ norm finite.

\subsection{Hard-sphere scattering}

Here we describe the architecture used to perform the hard-sphere scattering calculation of Section~\ref{sec:nss}. To compute $\tilde \psi_{\mathrm{out}}$, we construct a neural network which is defined only on the unit sphere. This takes as input the three coordinates $x_i / |x|$. This network is a fully-connected multi-layer perceptron (MLP). For the results of Section~\ref{sec:nss} we have three hidden layers, each of width $80$, using the ``swish'' activation function defined by
\begin{equation}
	\sigma_{\mathrm{swish}}(x) = \frac{x}{1 + e^{-x}}
	\text.
\end{equation}
This network has two outputs, which provide the real and imaginary parts of the asymptotic function $f$ which defines the outgoing wavefunction:
\begin{equation}
	\tilde \psi_{\mathrm{out}} = e^{-\left(\frac{1}{r}\right)^2} f \frac{e^{ikr}}r
	\text.
\end{equation}
The subleading terms in $\psi_{\mathrm{out}}$ are those which are required to make the $L_1$ norm finite.

The bulk wavefunction $\tilde \psi_{\mathrm{bulk}}$ is required to have asymptotic behavior such that $(E-\hat H)\tilde \psi_{\mathrm{bulk}}$ decays at least as $r^{-4}$. Therefore we write
\begin{equation}
	\tilde \psi_{\mathrm{bulk}}
	= \frac{e^{ikr}}{1 + r^3} \mathrm{NN}_{\mathrm{bulk}}
	\text.
\end{equation}
The bulk neural network $\mathrm{NN}_{\mathrm{bulk}}$ takes as input the full position vector without pre-normalization being applied. This MLP is chosen, in Section~\ref{sec:nss}, to also have $3$ hidden layers of width $80$. However, to ensure that the function $\mathrm{NN}_{\mathrm{bulk}}$ is asymptotically constant, the activation function on the final layer (after which exactly one linear transformation is applied) is taken to be the sigmoid:
\begin{equation}
	\sigma_{\mathrm{sigmoid}}(x) = \frac{1}{1 + e^{-x}}
	\text.
\end{equation}
To improve the learning ability of the bulk neural network, we feed in the radius $|x|$ in addition to the three positions $x_i$.

\subsection{Neutron-proton scattering}
In Section~\ref{sec:np} a similar ansatz is used as in the calculation of hard-sphere scattering, with only a few changes.

First, we parameterize the outgoing behavior $\tilde \psi_{\mathrm{out}}$ explicitly in terms of partial waves. This allows us to efficiently compute subleading behavior in $r^{-1}$ without finite differencing or automatic differentiation. We keep $\ell=0,1,2$ and set all other coefficients to zero. Given those partial wave amplitudes, the outgoing wavefunction is defined to include both $r^{-1}$ and $r^{-2}$ terms:
\begin{equation}
	\tilde \psi_{\mathrm{out}}
	= e^{-\left(\frac{r_{\mathrm{far}}}{r}\right)^2}\frac{e^{ikr}}{r} \left[
		\tilde f
		+
		\frac{\Delta \tilde f}{2i k r}
	\right]\text.
\end{equation}
Here $\tilde f$ is the sum of the partial waves up through $\ell=2$. The Laplacian is efficiently computed according to Appendix~\ref{app:laplacian}.

The bulk behavior is parameterized with a multi-layer perceptron with three inputs (the three relative coordinates), a width of $40$, and $4$ hidden layers. We do not enable a nonlinear final activation; intermediate activations are the ``swish'' function as before. In terms of this network the bulk behavior is given by
\begin{equation}
	\tilde \psi_{\mathrm{bulk}} =
	\frac{1}{1 + \left(\frac{r}{r_{\mathrm{near}}}\right)^4}
	\mathrm{NN}_{\mathrm{bulk}}
	\text.
\end{equation}
This approach relies on the fact that asymptotically, our multilayer perceptron grows at most linearly with the inputs.

The two parameters which define the decay of the bulk behavior, and onset of the asymptotic behavior, are initialized to be $r_{\mathrm{near}} = 3.0\,\mathrm{fm}$ and $r_{\mathrm{far}} = 0.5\,\mathrm{fm}$. We allow these parameters to be trained as part of the optimization.

\subsection{Neutron-deuteron scattering}

As described in the main text, in order to keep the ansatz architecture relatively simple, some modifications were performed to the loss function in the case of neutron-deuteron scattering, both above and below the deuteron breakup threshold.

In the case of elastic scattering (the calculations behind Figure~\ref{fig:nd-elastic}), only $s$-wave scattering is included in the ansatz, and therefore $\tilde\psi_{\mathrm{out}}$ is parameterized by a single complex number. We define a single ``window function'' as in Eq.~(\ref{eq:window}), and then the outgoing component is defined by Eq.~(\ref{eq:out-elastic}).

In the case of inelastic scattering (Figure~\ref{fig:nd-inelastic}), both the elastic and inelastic outgoing behaviors are parameterized by multi-layer perceptrons, of width $50$ and with one hidden layer on which $\sigma_{\mathrm{swish}}$ is applied. The elastic MLP takes $3$ inputs and yields $12$ (real) outputs; the inelastic MLP takes $6$ inputs and yields $16$ (real) outputs.

For the elastic scattering experiment the bulk behavior is parameterized by a two-layer MLP of width $250$, using $\sigma_{\mathrm{swish}}$ as the activation. The output of this MLP is scaled by the window function $w(r)$, described in the main text.

For inelastic scattering we take inspiration from the SIREN architecture~\cite{sitzmann2020implicit}. We again use an MLP (of depth $2$ for the smaller network, and $3$ for the larger), but the first layer has width of $2W$ (where $W$ is either $250$ or $500$), and the activation function used is $z \mapsto \sin \omega z$. We initialize the weights of this layer as described in~\cite{sitzmann2020implicit}. The parameter $\omega$ naturally has units of $\mathrm{fm}^{-1}$, and we take $\omega = 8\,\mathrm{fm}^{-1}$. Later hidden layers use $\sigma_{\mathrm{swish}}$ as the activation function, with a width of $W$ as before.

\section{The Laplacian on the hypersphere}\label{app:laplacian}
It is often necessary to have numerical access to the action of the Laplace-Beltrami operator on the (hyper)sphere. This can be achieved by automatic differentiation (or finite differencing) only at considerable cost, particularly when higher derivatives are required. In this appendix we outline an alternative approach which is particularly efficient in low dimensions and when the function on the sphere does not have high harmonic content.

We begin by noting that the Laplacian $\Delta$ on $\mathbb R^d$ can be decomposed into the spherical Laplacian $\Delta_\Omega$ on the $(d-1)$-sphere, and a radial term $\Delta_r$. This decomposition reads:
\begin{equation}\label{eq:laplacian}
	\Delta = r^{-2} \Delta_\Omega + \Delta_r\text.
\end{equation}
The radial term in the Laplacian has the same form in any number of dimensions, namely
\begin{equation}
	\Delta_r = r^{1-d}\frac{\partial}{\partial r} \left(r^{d-1} \frac{\partial}{\partial r}\right)
	\text.
\end{equation}

The $\mathbb R^d$ Laplacian is particularly efficient to evaluate on functions which are represented explicitly as polynomials of the Cartesian coordinates $x_1,\ldots,x_d$. For a general monomial $\prod_i x_i^{\alpha_i}$ the Laplacian is given by
\begin{equation}
	\Delta \prod_{i=1}^d x_i^{\alpha_i}
	=
	\sum_{j=1}^d \alpha_j (\alpha_j-1) x_j^{\alpha_j-2}
	\prod_{i \ne j} x_i^{\alpha_i}
	\text.
\end{equation}
The action of the Laplacian on polynomials is given by linearity. Similarly we may evaluate the radial term $\Delta_r$ on an arbitrary monomial by noting that $\frac{\partial}{\partial r} x_i = \frac{x_i}{r}$, and therefore the radial part of the Laplacian acting on any degree-$k$ homogeneous polynomial $P$ yields
\begin{equation}
	\Delta_r P = r^{-2} \left[(d-2)k + k^2\right] P
	\text.
\end{equation}
The radial part of the Laplacian does not yield a polynomial of the coordinates. However, when restricted to any sphere centered on the origin, the resulting expression is equal to a polynomial. Here we restrict to the surface $\sum_i x_i^2 = 1$, corresponding to setting $r=1$ throughout. On this surface, the radial Laplacian is equal to
\begin{equation}
	\Delta_r P = \left[k^2 + k(d-2)\right]P\text,
\end{equation}
where again $P$ is a homogeneous polynomial of degree $k$.

Thus, both $\Delta$ and $\Delta_r$, when restricted to the sphere, may be efficiently written as linear operators acting on the vector space spanned by monomials of the Cartesian coordinates. These operators do not increase the degree of the monomials, meaning that if we truncated this space to the vector space of polynomials of degree $\le \nu$, both $\Delta$ and $\Delta_r$ are still equal to matrices on this truncated space. Using Eq.~(\ref{eq:laplacian}), the same is true of the Laplace-Beltrami operator $\Delta_\Omega$ on any $S_{d-1}$.

\section{Approximated outgoing asymptotics}\label{app:outgoing}
If we write an ansatz for a scattering state (for concreteness, in three-dimensional elastic scattering from a central potential) of the form
\begin{equation}
	\tilde\psi = \psi_{\mathrm{in}} + \frac{e^{ikr}}{r} f + O(e^{-r})\text,
\end{equation}
then for nontrivial functions $f$, the $L_1$ norm of the Schr\"odinger violation is divergent. In Section~\ref{sec:np}, we dealt with this by including the next-to-leading term in $r^{-1}$, at some computational expense. In Section~\ref{ssec:elastic}, we limited the calculation to $s$-wave scattering, in which the next-to-leading term vanishes exactly.
In Section~\ref{ssec:inelastic}, an approximated loss function was constructed which did not require any subleading terms to be computed. This approximation, while controlled, was not systematically improvable.

The purpose of this appendix is to describe a method by which an approximated loss function $\tilde {\mathcal L}$ may be constructed which, in an appropriate and computationally accessible limit, becomes an arbtirarily good approximation for the ``ideal'' loss function $\mathcal L$. This follows the discussion in Section~\ref{ssec:inelastic}.

In addition to the leading-order function $\tilde f^{(0)}$, we parameterize $m$ functions $\tilde g^{(1)},\ldots,\tilde g^{(m)}$ by neural networks. Defining $g^{(0)} = f^{(0)}$, these together define $\tilde G$ by a finite power series:
\begin{equation}
	\tilde G =  r^{\frac{1-d}{2}} \sum_{j=0}^m r^{-j} \tilde g^{(j)}
	\text.
\end{equation}
These functions $g^{(j)}$ may be trained at reasonable computational cost by building a loss function from Eq.~(\ref{eq:f-recursion}). For example we may define
a loss function $\mathcal K$ by
\begin{widetext}
\begin{equation}
	\mathcal K[\tilde g] =
	\sum_{j=0}^{m-1} \int \left|
	-2ik(j+1) \tilde g^{(j+1)} + \Delta_\Omega \tilde g^{(j)}
	- \left(\frac{d-1}{2} + j\right)\left(\frac {d-1} 2 - j - 1\right) \tilde g^{(j)}
	\right|
	\text,
\end{equation}
\end{widetext}
the evaluation of which does not require any nested Laplacians, no matter what value is taken for $m$.

Because the sum in $\mathcal K$ is finite, even at $\mathcal K = 0$ there is still some error introduced in the approximated loss function of Eq.~(\ref{eq:loss-approx-outgoing}). We could improve that error by increasing $m$, but at considerable computational cost due to the need to store and train more neural networks. However, if $m$ is large enough, the Schr\"odinger violation decays fast enough at large radii that we can also improve the approximation by taking the parameter $r_0$ defining the window function $w(r)$ to be larger.

\bibliographystyle{apsrev4-2}
\bibliography{refs}

\end{document}